\documentclass[conference,compsoc]{IEEEtran}
\ifCLASSINFOpdf
\else
\fi
\usepackage{url}

\usepackage{amsfonts,amsthm}
\usepackage{array}
\usepackage{booktabs}
\usepackage{caption}
\usepackage{color}
\usepackage{enumitem}
\usepackage{float}
\usepackage{footnote}
\usepackage{mathtools}
\usepackage{makecell}
\usepackage{multirow}
\usepackage{rotating}
\usepackage{setspace}
\usepackage{subcaption}

\theoremstyle{definition}

\newcolumntype{L}[1]{>{\raggedright\let\newline\\\arraybackslash\hspace{0pt}}m{#1}}
\newcolumntype{C}[1]{>{\centering\let\newline\\\arraybackslash\hspace{0pt}}m{#1}}
\newcolumntype{R}[1]{>{\raggedleft\let\newline\\\arraybackslash\hspace{0pt}}m{#1}}

\setlist[itemize]{noitemsep, topsep=0pt}
\setlist[enumerate]{noitemsep, topsep=0pt}

\newcommand{\sota}{state-of-the-art}

\newcommand{\parens}[1]{\left(#1\right)}
\newcommand{\braces}[1]{\left\{#1\right\}}
\newcommand{\bracks}[1]{\left[#1\right]}

\newcommand{\norm}[1]{\left\Vert#1\right\Vert}

\DeclarePairedDelimiter{\ceil}{\lceil}{\rceil}

\begin{document}
%
\title{Speech2AffectiveGestures: Synthesizing Co-Speech Gestures with Generative Adversarial Affective Expression Learning}

\author{
\IEEEauthorblockN{Uttaran Bhattacharya}
\IEEEauthorblockA{University of Maryland\\College Park, MD, USA\\
\url{uttaranb@umd.edu}}
\and
\IEEEauthorblockN{Elizabeth Childs}
\IEEEauthorblockA{University of Maryland\\College Park, MD, USA\\
\url{ehchilds@terpmail.umd.edu}}
\and
\IEEEauthorblockN{Nicholas Rewkowski}
\IEEEauthorblockA{University of Maryland\\College Park, MD, USA\\
\url{nick1@.umd.edu}}
\and
\IEEEauthorblockN{Dinesh Manocha}
\IEEEauthorblockA{University of Maryland\\College Park, MD, USA\\
\url{dmanocha@.umd.edu}}
}


%


\twocolumn[{%
\renewcommand\twocolumn[1][]{#1}%
\maketitle
\begin{center}
    \centering
    \captionsetup{type=figure}
    \includegraphics[width=\textwidth,height=5cm]{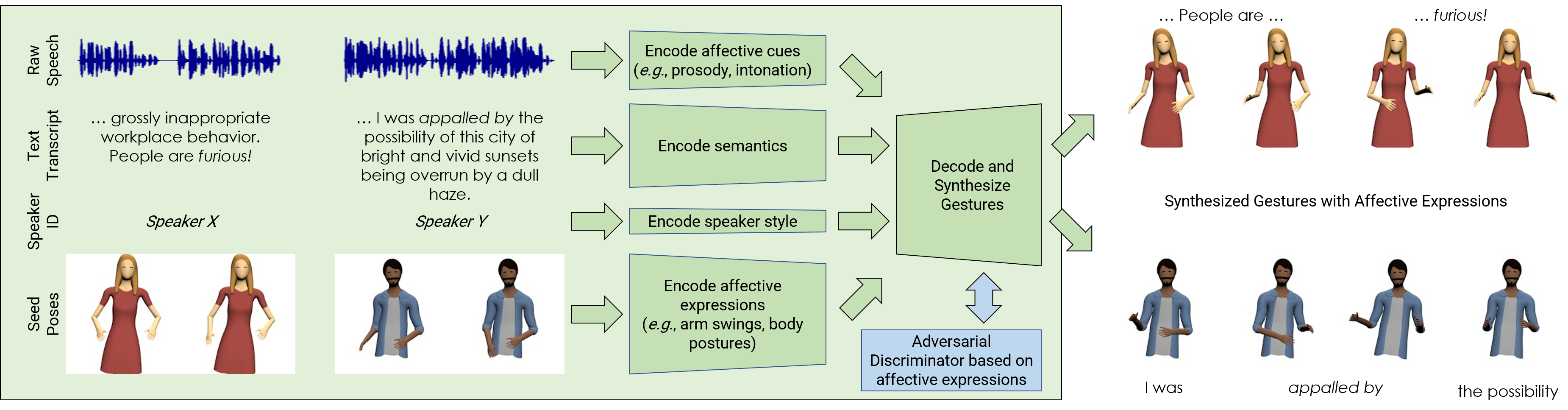}
    \captionof{figure}{We synthesize 3D pose sequences of co-speech upper-body gestures with appropriate affective expressions. We extract the affective cues from the speech, the sentiments from the corresponding text transcripts, the individual speaker styles, and the joint-based affective expressions from the seed poses (shown on the left). We train a generative adversarial network to synthesize gestures aligned with the speech by leveraging the affective information in both the generation and the discrimination phases. We show two such affective gestures on the right, with the affects \textit{furious} and \textit{appalled} denoted in italics.}
    \label{fig:teaser}
\end{center}%
}]

\begin{abstract}
We present a generative adversarial network to synthesize 3D pose sequences of co-speech upper-body gestures with appropriate affective expressions. Our network consists of two components: a generator to synthesize gestures from a joint embedding space of features encoded from the input speech and the seed poses, and a discriminator to distinguish between the synthesized pose sequences and real 3D pose sequences. We leverage the Mel-frequency cepstral coefficients and the text transcript computed from the input speech in separate encoders in our generator to learn the desired sentiments and the associated affective cues. We design an affective encoder using multi-scale spatial-temporal graph convolutions to transform 3D pose sequences into latent, pose-based affective features. We use our affective encoder in both our generator, where it learns affective features from the seed poses to guide the gesture synthesis, and our discriminator, where it enforces the synthesized gestures to contain the appropriate affective expressions. We perform extensive evaluations on two benchmark datasets for gesture synthesis from the speech, the TED Gesture Dataset and the GENEA Challenge 2020 Dataset. Compared to the best baselines, we improve the mean absolute joint error by 10--33\%, the mean acceleration difference by 8--58\%, and the Fr\'echet Gesture Distance by 21--34\%. We also conduct a user study and observe that compared to the best current baselines, around 15.28\% of participants indicated our synthesized gestures appear more plausible, and around 16.32\% of participants felt the gestures had more appropriate affective expressions aligned with the speech.
\end{abstract}



%

\section{Introduction}\label{sec:intro}
Co-speech gestures are bodily expressions associated with a person's speech~\cite{cospeech_gestures}. They help underline the subject matter and the context of the speech, particularly in the form of beat, deictic, iconic, or metaphoric expressions~\cite{gestures_in_comm}. Beat gestures are rhythmic movements following the speech, and deictic gestures point to an entity. Iconic gestures describe physical concepts, \textit{e.g.}, spreading and contracting the arms to denote ``large'' and ``small'', and metaphoric gestures describe abstract concepts, \textit{e.g.}, putting a hand to the heart to denote ``love''. Synthesizing co-speech gestures is an important task in creating socially engaging characters and virtual agents. These are useful in a variety of multimedia application such as online learning~\cite{online_learning1,online_learning2,online_learning3}, interviewing and counseling~\cite{interviewing,simsensei}, robot assistants~\cite{cospeech_gestures}, character designs and game development~\cite{game_development,gesticulator}, and visualizing stories and scripts~\cite{script_visualization}.

In our work, we focus on synthesizing the upper-body gestures associated with speech. We consider the joints at the root, spine, head, and the two arms as part of the upper body, which are the joints most commonly used in co-speech gestures~\cite{trimodal,gesticulator}. Current \sota~methods for co-speech upper-body gesture synthesis are based on an end-to-end learning approach~\cite{individual_gesture_styles,trimodal,gesticulator}. These methods train deep neural networks using gestures (available as videos or motion-captured datasets), raw speech waveforms and the corresponding text transcripts, and individual speaker styles. While these methods can generate different beat, deictic, iconic, and metaphoric co-speech gestures and adapt to speaker-specific styles, they do not have any mechanism to reliably incorporate affective expressions in the gestures.

Affective expressions are the modulations in gestures resulting from the emotions experienced by the speakers~\cite{affective_body_survey1,affective_body_survey2}. Even for a given speaker, the style of gesture expressions can change depending on the emotional context, and human observers are keenly alert to these changes~\cite{affective_body_survey2}. The combined understanding of the content of the speech and the speaker's gesture-based affective expressions are crucial to human-human interactions~\cite{human_human1,human_human2}. Therefore, it is essential to incorporate affective expressions in co-speech gestures of animated characters and virtual agents to improve their plausibility in human-machine interactions.

In human-human interactions, we can break the gesture-based affective expressions down into a set of biomechanical features known as \textit{affective features}, such as body postures, head positions, and arm motions~\cite{affective_body_survey1}. Each affective expression is a combination of one or more affective features, \textit{e.g.}, rapid arm swings and head jerks are often used as expressions of anger or excitement~\cite{affective_body_survey2}. A multitude of macroscopic and microscopic factors influence the affective features in a given context, including the social setting and the speaker's idiosyncrasies, making an exhaustive enumeration of affective features tedious and challenging~\cite{taew}. Nevertheless, it is essential to learn these affective features to understand and synthesize the desired affective expressions.

Moreover, co-speech affective gesture synthesis also requires aligning the gestures with the affective cues obtained from the speech. To this end, prior methods have either learned to map the raw speech waveforms to gestures via latent embeddings~\cite{trimodal} or utilized the log-Mel spectrograms to obtain a richer understanding of the affective cues, including the prosody and the intonations in the speech~\cite{individual_gesture_styles,gesticulator}. However, these are high-dimensional representations of speech that require significant computation overhead to be downscaled into convenient latent embedding spaces.

\noindent\textbf{Main Contributions.} We present an end-to-end learning approach for generating 3D pose sequences of co-speech gestures with appropriate affective expressions while maintaining the speakers' individual styles and following a short sequence of seed poses.\footnote{Code and additional materials available at \url{https://gamma.umd.edu/s2ag}.} 
We leverage the Mel-frequency cepstral coefficients (MFCCs) from the speeches obtained by performing DCT on the log-Mel spectrograms. MFCCs are highly compressible representations containing sufficient information for speaker identification and also encode affective cues such as prosody and intonation for speech-based emotion recognition. We use separate encoders to encode the MFCCs from the raw speeches, the text transcripts obtained from the speeches, the speakers' styles, and the seed poses. We use available text- and speaker-encoders proposed by Yoon et al.~\cite{trimodal} to learn latent features from the text transcript and a latent style embedding space using a variational encoding of the speaker styles. We propose an encoder for the MFCCs that captures the affective cues in the speech. We also develop an ``affective encoder'' that transforms the 3D pose sequences to latent affective features using multi-scale spatial-temporal graph convolutions (STGCNs).
We design our multi-scale STGCNs to expand attention from the local joints to the macroscopic body parts in a bottom-up manner. We use our affective encoder both in the generator to learn affective features from the seed poses to guide the gesture synthesis and in our discriminator to differentiate between the real and the synthesized gestures based on the affective expressions. To the best of our knowledge, we are the first to learn affective features directly from the gesture data to synthesize gestures with affective expressions. Our main contributions include:

\begin{itemize}
    \item \textbf{Synthesizing co-speech affective gestures.} We synthesize 3D pose sequences of gestures with appropriate affective expressions given a speaker's speech, maintaining the speakers' individual styles of gesticulation and following a short sequence of seed poses.

    \item \textbf{Affective encoder for learning latent affective features.} Our affective encoder leverages the localized joint movements and the macroscopic body movements in the 3D pose sequences to learn latent affective features that are used for synthesizing the future poses from the seed poses and adversarially guiding the synthesis as per affective expressions.
    
    \item \textbf{MFCC encoder for leveraging the affective cues from the speech.} Our MFCC encoder takes in low-dimensional MFCCs containing information on the affective cues from the speech, including prosody and intonations, and transforms them into latent embeddings for affective gesture synthesis.
\end{itemize}

We evaluate the quantitative performance of our network on two benchmark datasets, the TED Gesture Dataset~\cite{cospeech_gestures} and the GENEA Challenge 2020 Dataset~\cite{genea_2021}. We observe an improvement of 10--33\% on the mean absolute joint error, 8--58\% on the mean acceleration difference, and 21--34\% on the Fr\'echet Gesture Distance (FGD)~\cite{trimodal} for our network compared to the current state-of-the-art baselines. We also conduct a user study to evaluate the plausibility of our synthesized gestures and the consistency between the affective expressions in the gestures and the speech. Around 15.28\% participants indicated that our synthesized gestures are more plausible than the best current baseline of Yoon et al.~\cite{trimodal}, and around 16.32\% participants felt the gestures had more appropriate affective expressions aligned with the speech compared to the same baseline.

\section{Related Work}\label{sec:rw}
We briefly summarize related prior work on how humans perceive affective body expressions and how these studies were leveraged to synthesize emotionally expressive characters. We also summarize works on synthesizing body motions, especially those aligned with a speech, a text transcript, or both.

\subsection{Perceiving Affective Body Expressions}
Affect is traditionally expressed in psychology in terms of its valence, arousal, and dominance (VAD)~\cite{vad}. Valence measures the level of pleasantness (\textit{e.g.}, happy vs. sad), arousal measures how animated the person is (\textit{e.g.}, angry vs. bored), and dominance measures the level of control over the affect (\textit{e.g.}, admiration vs. fear). Studies in both psychology and affective computing indicate the existence of biomechanical \textit{affective} features that provide cues to a person's perceived affect to human observers~\cite{affective_body_survey1,affective_body_survey2,emotion_in_gesture,learning_unseen_emotions,kim2015brvo}. These affective features can be observed at different scales: they can be localized joint movements such as rapid arm swings and head jerks, indicating excitement or anger, as well as macroscopic body movements such as the upper body being expanded, indicating pride or confidence, or collapsed, indicating shame or nervousness. Subsequently, there has been work on detecting perceived emotions by leveraging known affective features either as input to a neural network~\cite{step} or to constrain the embedding space~\cite{taew}. In contrast, we design our neural network to explicitly attend to the body movements at these multiple scales to learn latent affective features directly from the input gesture samples.

\subsection{Synthesizing Affective Body Expressions}
There has been substantial work on synthesizing affective expressions for embodied conversation agents~\cite{verbal_comm1,verbal_comm2} and other social virtual agents to interact via facial expressions~\cite{expressive_face1,expressive_face2} or gaits~\cite{eva}. Furthermore, the synthesis of affective facial expressions has been aligned with a character's speech using data-driven techniques~\cite{automated_emotive_agents}. While synthesizing speech-aligned affective facial expressions has been relatively well-studied, aligning the speech with affective body expressions has been more challenging. Some of the widely used approaches are rule-based systems such as that of DeVault et al.~\cite{simsensei}, which has a virtual human counselor expressing appropriate affective hand and body gestures following known mappings between the emotional states and the stored animations. Recent methods utilize gait datasets annotated with categorical emotions such as happy, sad, and angry to generate emotive gaits~\cite{tanmay_emotions,gen_emo_gaits}. Other techniques have extended to the VAD space of affect, where body gestures are generated given the text transcripts of speech and the corresponding intended emotion as a point in the VAD space~\cite{t2g}. Our approach is based on designing an end-to-end system that can synthesize body expressions by automatically understanding the affective content in the input speech.

\subsection{Synthesizing Gestures}
There is a rich body of work gesture synthesis using rule-based systems, as surveyed comprehensively by Wagner et al.~\cite{rule_based_gestures}. However, scalability to novel scenarios remains a challenge for rule-based systems on account of manually designing new rules. Instead, we focus on a summary of the recent data-driven approaches of automated gesture synthesis in novel scenarios~\cite{dcnf}, which are in line with our learning-based approach. Existing techniques have utilized hidden Markov models~\cite{speech_hmm_gestures}, recurrent neural network variants~\cite{speech_LSTM_gestures,cospeech_gestures}, and autoencoders~\cite{speech_RepL_gestures} to learn robust latent features that encode the input speech, available as either an audio or a text transcript, and can be used to decode the output gestures. Other approaches have opted to learn stochastic generation processes using tools such as invertible sub-transformations~\cite{speech_flow_gestures} to map between the speech and the gesture spaces. To improve the realism of the generated speech-driven gestures, more recent works incorporate the speech semantics into the training process~\cite{gesticulator}, and even combined the synthesized gestures with rule-based head nods and hand waves for embodied conversation agents~\cite{prototypical_behaviors}.

Our approach is complementary to these approaches in that we learn mappings from the text transcripts of speech to gestures. It eliminates the noise in speech signals and helps us focus only on the relevant content and context. Learning from the text also enables us to focus on a broader range of gestures, including iconic, deictic, and metaphoric gestures~\cite{gestures_in_comm}. Our work is most closely related to that of Yoon et al.~\cite{cospeech_gestures}. They learn upper body gestures as PCA-based, low-dimensional pose features, corresponding to text transcripts from a dataset of TED-talk videos, then map these 3D gestures to an NAO robot. They have also followed up this work by generating upper-body gestures aligned with the three modalities of speech, text transcripts, and person identity~\cite{trimodal}. On the other hand, we learn to map text transcripts to 3D pose sequences corresponding to semantic-aware, full-body gestures of more human-like virtual agents using an end-to-end trainable transformer network and blend in emotional expressiveness.

\subsection{Incorporating Speaker Styles}
Co-speech gesture generation is intrinsically related to stylized gesture generation. There has been considerable progress on stylized generation of head motions~\cite{head_motion_ae,head_gan}, facial motions~\cite{obamanet,expressive_face2} as well as locomotions~\cite{motion_editing,pfnn,nsm}. At the same time, many techniques have been proposed to generate appropriately styled body motions from textual descriptions of the actions~\cite{video_from_text,text_guided_pose}. Other approaches have developed separate gesture generation networks for individual speakers to adapt to their individual styles~\cite{individual_gesture_styles}, together with adversarial losses to improve the fidelity of the generation~\cite{multi_adversarial_gestures}. Recently, Yoon et al.~\cite{trimodal} proposed a unified architecture that considers the speech, its text transcript, and the speaker identity to generate co-speech gestures with continuously varying speaker styles. We extend such speaker-aware gesture synthesis to further incorporate the appropriate affective body expressions that align with the affective content in the speech.

\begin{figure}[t]
    \centering
    \includegraphics[width=\columnwidth]{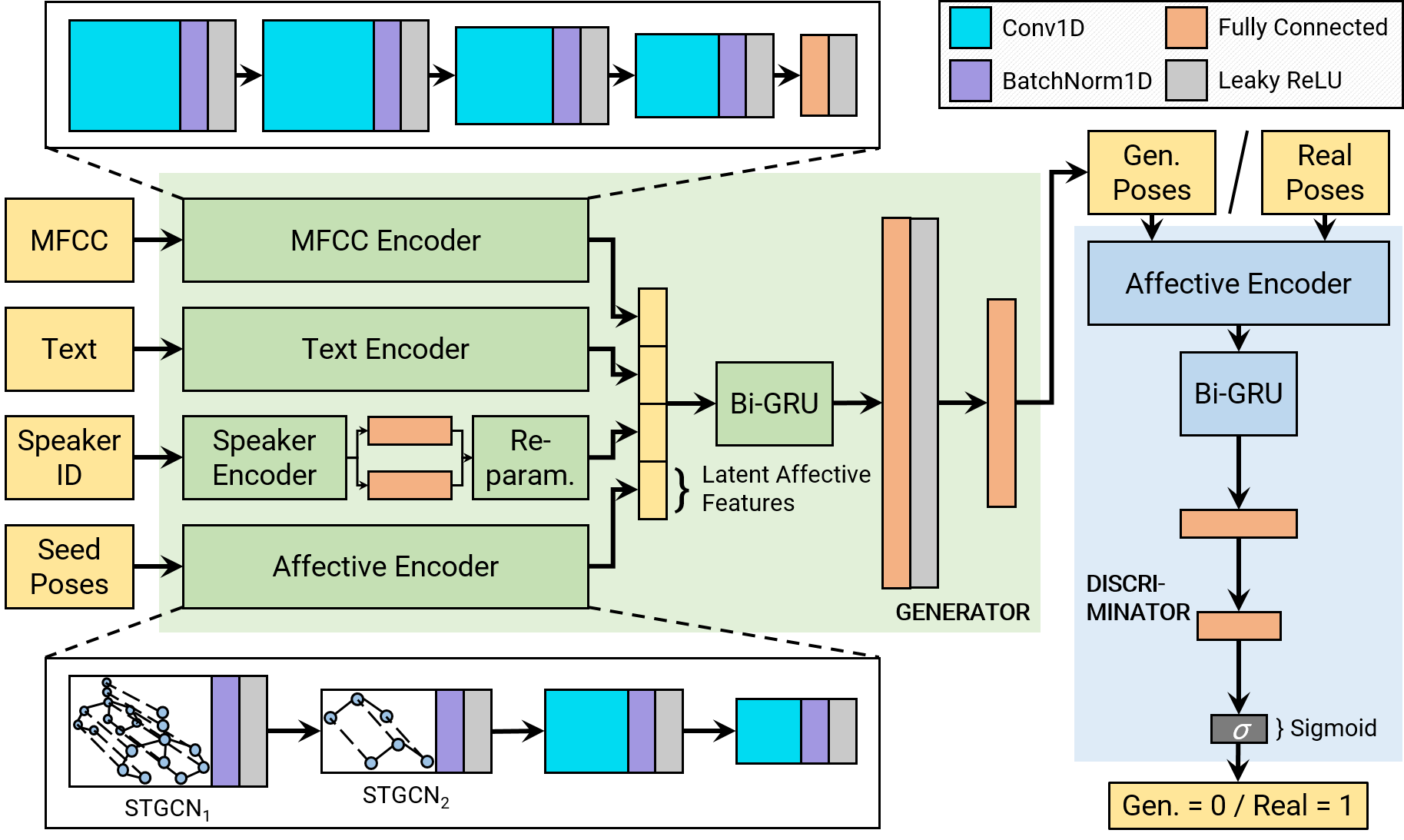}
    \caption{Our network consists of a generator (pale-green box) and a discriminator (pale-blue box). Our generator takes in the MFCC from the speech, the text transcript, the speaker ID, and a sequence of 3D seed poses. We use four encoders: the MFCC encoder (Sec.~\ref{subsubsec:mfcc_encoder}), the text encoder (Sec.~\ref{subsubsec:text_encoder}), the speaker encoder (Sec.~\ref{subsubsec:speaker_encoder}), and the affective encoder (Sec.~\ref{subsubsec:aff_encoder}). We feed the concatenation of these latent features into our Bi-GRU followed by a set of FC layers to synthesize the gestures aligned with the speech. Our discriminator learns to discriminate between the real and the synthesized gestures based on the latent affective features from the affective encoder, constraining the generator to synthesize appropriate affective expressions.}
    \label{fig:s2ag_network}
    \vspace{-10pt}
\end{figure}

\section{Approach}\label{sec:approach}
Our goal is to generate 3D pose sequences of co-speech upper-body gestures with appropriate affective expressions and speaker styles, given the raw speech waveform, the speaker identity, and a short sequence of seed poses. We consider affective expressions to be specific sequences of joint movements, generally as a combination of the affective features~\cite{t2g}. We learn these affective expressions both at the localized joint neighborhoods and the macroscopic body movements, and use them to condition the training of a generative adversarial network. We show our overall network architecture in Fig.~\ref{fig:s2ag_network}.

\subsection{Synthesizing Co-Speech Gestures}
Our generative network takes in the raw speech waveform as a 1D array, the corresponding text transcript as a sequence of words, the speaker identity as a unique number, and the seed poses as a 3D pose sequence. Similar to Yoon et al.~\cite{trimodal}, we encode the speech waveform, the text transcript, and the speaker identities using separate encoders. However, unlike Yoon et al., we convert the speech waveform to Mel-Frequency Cepstral Coefficients (MFCCs) to guide the encoding process based on the affective cues from speech. We also propose an affective encoder to encode the pose-based affective expressions into latent features for both gesture generation and discrimination. In the generation process, we combine the latent embeddings learned from the four encoders, speech, text, speaker, and affective, into a joint embedding for learning the upper-body gestures.

\subsubsection{MFCC Encoder}\label{subsubsec:mfcc_encoder}
MFCCs are known to encode signal frequencies consistent with how humans perceive sound, and are therefore particularly useful for tasks such as speech recognition~\cite{mfcc_speech_recognition}, speaker identification~\cite{mfcc_speaker_identification} and speech-based emotion recognition~\cite{mfcc_ser}. In our case, we design our MFCC encoder to embed the speech-based affective cues such as prosody and intonations captured by the MFCCs and incorporate them in gesture synthesis. Given a raw waveform as a 1D array, we transform it to its top 14 MFCCs. These include the log-energy spectrum and 13 coefficients containing sufficient information on the speaker's pitch, intonation, prosody, and other relevant parameters~\cite{cmu_mosei,obamanet}. We also append the first- and the second-order discrete forward differences of the 13 coefficients, obtaining a total of 37 values. Using a window size $W$ on a input waveform of length $L$, we obtain individual MFCCs of shape $\ceil*{L/W}$, leading to a combined feature tensor $f_m \in \mathbb{R}^{37 \times \ceil*{L/W}}$. We pass these features through a series of 1D temporal convolutions, followed by a single fully-connected (FC) layer, to obtain a latent feature sequence $\hat{f}_m \in \mathbb{R}^{D_m \times T}$ of sequence length $T$ equal to 3D pose sequence length of the seed poses, as
\begin{equation}
    \hat{f}_m = \textrm{Conv}\circ\textrm{FC}_{mfcc}\parens{f_m; W_{mfcc}},
\end{equation}
where $D_m$ is the dimension of the latent features, $\textrm{Conv}\circ\textrm{FC}_{mfcc}$ denotes the series of 1D convolutions followed by the FC layer, and $W_{mfcc}$ its set of trainable parameters.

\subsubsection{Text Encoder}\label{subsubsec:text_encoder}
Given the text transcript corresponding to the speech, we first pad the transcript with padding tokens following the approach of Yoon et al., to ensure that the text transcript has the same sequence length $T$ as the seed poses. We then use the pre-trained FastText~\cite{fasttext} word embedding model to transform the word sequence into 300-dimensional features, leading to a feature tensor $f_x \in \mathbb{R}^{300 \times T}$. We use FastText for its memory efficiency and its usefulness in sentiment analysis~\cite{fasttext_sentiment_analysis}, which is important in understanding text-based affect. We pass the FastText features through a series of temporal 1D convolutions to obtain a latent feature sequence $\hat{f}_x \in \mathbb{R}^{D_x \times T}$, as
\begin{equation}
    \hat{f}_x = \textrm{Conv}_{text}\parens{f_x; W_{text}},
\end{equation}
where $D_x$ is the dimension of the latent features, $\textrm{Conv}_{text}$ denotes the series of 1D convolutions with trainable parameters $W_{text}$.

\subsubsection{Speaker Encoder}\label{subsubsec:speaker_encoder}
For the speaker IDs, we use one-hot vectors $f_s \in \braces{0, 1}^\mathcal{S}$, assuming $\mathcal{S}$ is the number of available speakers. Following Yoon et al.~\cite{trimodal}, we use two sets of FC layers to learn an embedding space capturing the mean $\mu_s \in \mathbb{R}^{D_s}$ and the variance $\Sigma_s \in \mathbb{R}^{D_s \times D_s}_+$ of the latent distribution of the speaker styles as
\begin{align}
    \mu_s &= \textrm{FC}_\mu\parens{f_s; w_\mu} \\
    \log\Sigma_s &= \textrm{FC}_\Sigma\parens{f_s; w_\Sigma},
\end{align}
where $D_s$ is the dimension of the latent distribution space, $\textrm{FC}_\mu$ and $\textrm{FC}_\Sigma$ denote the two sets of FC layers, and $W_\mu$ and $W_\Sigma$ denote the corresponding sets of trainable parameters. Intuitively, this latent distribution space consists of all the available speakers, plus speakers that can be ``constructed'' by linear combinations of those speakers in the latent space. As a result, we can pick a random point from the latent space to use in the synthesis, resulting in some variability in the synthesized gestures even when the speech remains the same. We term this variability as having ``speaker-aware'' styles. Given the parameters $\mu_s$ and $\Sigma_s$ of latent distribution space, we use the re-parametrization trick~\cite{vae} to generate a random speaker-aware style sample $\hat{f}_s \in \mathbb{R}^{D_s}$ and repeat it for all the $T$ time steps of the input pose sequence. 

\subsubsection{Affective Encoder}\label{subsubsec:aff_encoder}
We propose an encoding mechanism that transforms the pose-based affective expressions into a latent embedding. Since gestures typically consist of movements in the trunk, arms, and head, we only consider ten joints corresponding to these parts of the body: root, spine, neck, head, left and right shoulders, left and right elbows, and left and right wrists. We consider a directed graph for the pose, where the joints are the vertices, and the edges are directed from the root towards the extremities.
We assume the edge lengths are known for each input and train our encoder only on the directions of the edges. We consider nine unit-vector sequences $U = \bracks{u_1; \dots; u_9}$, each of sequence length $T$, to denote the edge directions at the corresponding $T$ time steps of the input pose sequence.

We employ a hierarchical encoding strategy using spatial-temporal graph convolutions (STGCNs)~\cite{stgcn}. STGCNs are adapted to leverage localized dependencies in generalized graph-structured data, and are therefore suitable for our pose graph sequences. We use two levels of hierarchy, the first at the level of individual bones and the second at the level of the three body parts, the trunk and the two arms. At the first level, our unweighted adjacency matrix $A_1 \in \braces{0, 1}^{9 \times 9 \times T}$ captures the temporal counterparts of each edge at the four nearest time steps (past two and future two), and spatially adjacent edges with a maximum hop of two, \textit{i.e.}, we consider two edges to be spatially adjacent if they either share a vertex or are connected to the two ends of a third edge. This size of the adjacent neighborhood sufficiently groups the edges influenced by typical affective expressions such as arm swings, head jerks, and upper-body collapse. Consequently, the convolution filters can learn a latent feature sequence $\hat{f}_{a_1} \in \mathbb{R}^{D_{a_1} \times 9 \times T}$ from the edges based on the variations in the affective expressions, obtained as
\begin{equation}
    \hat{f}_{a_1} = \textrm{STGCN}_1\parens{U, A_1; W_{a_1}},
\end{equation}
where $D_{a_1}$ is the dimension of the per-edge latent features, $\textrm{STGCN}_1$ denotes the first-level STGCN with trainable parameters $W_{a_1}$. At the second level, the three body parts, the trunk and the two arms, capture the macroscopic body movements such as raising or crossing the arms, and bending or straightening the trunk. In the second-level adjacency matrix $A_2 \in \braces{0, 1}^{3 \times 3 \times T}$, we assume both the arms to be adjacent to the torso but not to each other, since the movements on one arm need not influence the other. We again consider the temporal counterparts of each body part in the four nearest time steps in the temporal adjacency. We reshape the latent features $\hat{f}_{a_1}$ to $3D_{a_1} \times 3 \times T$, to collect the per-edge features corresponding to the three body parts in the feature dimension. Our second-level STGCN then operates on these reshaped features to produce the second-level latent features $\hat{f}_{a_2} \in \mathbb{R}^{D_{a_2} \times 3 \times T}$ as
\begin{equation}
    \hat{f}_{a_2} = \textrm{STGCN}_2\parens{\hat{f}_{a_1}, A_2; W_{a_2}},
\end{equation}
where $D_{a_2}$ is the dimension of the per-edge latent features, $\textrm{STGCN}_2$ denotes the second-level STGCN with trainable parameters $W_{a_2}$. We then apply a series of 1D convolutions on the reshaped second-level features $\hat{f}_{a_2} \in \mathbb{R}^{3D_{a_2} \times T}$ to obtain the latent affective feature sequence $\hat{f}_a \in \mathbb{R}^{D_a \times T}$, as
\begin{equation}
    \hat{f}_a = \textrm{Conv}_{aff}\parens{\hat{f}_{a_2}; W_{a}},
\end{equation}
where $D_a$ is the dimension of the latent affective features, and $\textrm{Conv}_{aff}$ denotes the series of 1D convolutions with trainable parameters $W_{a}$.

\subsubsection{Gesture Generator}
Given the latent feature sequences $\hat{f}_m$, $\hat{f}_x$, $\hat{f}_s$, and $\hat{f}_a$, we concatenate them, pass them through a bidirectional gated recurrent unit (Bi-GRU), and sum the bidirectional outputs to obtain the predicted edge embeddings sequence $\hat{u}_e \in \mathbb{R}^{D_{e} \times T}$, as
\begin{align}
    out_{frw}, out_{bkw} &= \textrm{GRU}_e\parens{\bracks{\hat{f}_m; \hat{f}_x; \hat{f}_s; \hat{f}_a}; W_e}, \\
    \hat{u}_e &= out_{frw} + out_{bkw},
\end{align}
where $D_e$ is the dimension of the predicted edge embeddings, $\textrm{GRU}_e$ denotes the bidirectional GRU with the corresponding set of trainable parameters $W_e$, and $out_{frw}$ and $out_{bkw}$ respectively denote the outputs of the forward and the backward channels of the GRU. As in Yoon et al., we then transform the predicted edge embeddings to predicted edge vector sequences $\hat{U} = \bracks{\hat{u}_1; \dots; \hat{u}_9}$, each of sequence length $T$, using a set of FC layers as
\begin{equation}
    \hat{U} = \textrm{FC}_{gen}\parens{\hat{u}_e; W_{gen}},
\end{equation}
where $\textrm{FC}_{gen}$ denotes the set of FC layers with the trainable parameters $W_{gen}$. Thus, our generator is designed to take in a sequence of seed poses of length $T$ and predicts a pose sequence of gestures for the next $T$ time steps. Finally, we scale each predicted edge vector $\hat{u}_i$ to have the corresponding bone length $b_i$, $i = 1, \dots, 9$. We add it to the 3D position $pos_{s\parens{i}}$ of the source joint $s\parens{i}$ of that edge vector to obtain 3D position $pos_{d\parens{i}}$ of the destination joint $d\parens{i}$ of the same edge vector, as
\begin{equation}
    pos_{d\parens{i}} = pos_{s\parens{i}} + b_i\cdot\frac{\hat{u}_i}{\norm{\hat{u}_i}}.
\end{equation}

\subsection{Discriminating Gestures}\label{subsec:discriminator}
Our discriminator takes in a gesture of sequence length $T$ and computes its latent affective feature sequence $\hat{f}_a \in \mathbb{R}^{D_a \times T}$ using our affective encoder (Sec.~\ref{subsubsec:aff_encoder}). We pass this feature sequence through another bidirectional GRU, and sum the bidirectional outputs to obtain the discriminator embeddings sequence $\hat{d} \in \mathbb{R}^{h \times T}$, as
\begin{align}
    out_{frw}, out_{bkw} &= \textrm{GRU}_{disc}\parens{\hat{f}_a; W_{GRU, disc}}, \\
    \hat{d} &= out_{frw} + out_{bkw},
\end{align}
where $D_d$ is the dimension of the predicted discriminator embeddings, $\textrm{GRU}_{disc}$ denotes the bidirectional GRU with trainable parameters $W_{GRU_disc}$, and $out_{frw}$ and $out_{bkw}$ respectively denote the outputs of the forward and the backward channels of the GRU. We then transform the discriminator embeddings to a probability vector $c \in \bracks{0, 1}$ using a set of FC layers as
\begin{equation}
    c = \textrm{FC}_{disc}\parens{\hat{d}; W_{FC disc}},
\end{equation}
where $\textrm{FC}_{disc}$ denotes the set of FC layers with trainable parameters $W_{FC disc}$, and $c$ is such that $c \geq 0.5$ implies the discriminator predicts the input gesture to be real, and generated otherwise.

\begin{table}[t]
    \centering
    \caption{Hyperparameters (HPs) for our network. We chose all the values via empirical search.}
    \label{tab:hyperparams}
    \resizebox{\columnwidth}{!}{
    \begin{tabular}{rlr}
        \toprule
        HP & Description & Value \\
        \midrule
        $D_m$ & Latent feature from the MFCC encoder & 32  \\
        $D_x$ & Latent feature from the text encoder & 32  \\
        $D_s$ & Latent distribution space of speaker styles & 16  \\
        $D_{a_1}$ & Per-edge latent features after $\textrm{STGCN}_1$ in the affective encoder & 16 \\
        $D_{a_2}$ & Per-edge latent features after $\textrm{STGCN}_2$ in the affective encoder & 16 \\
        $D_a$ & Latent affective features from the affective encoder & 16 \\
        $D_e$ & Predicted edge embeddings from the GRU in the generator & 150 \\
        $D_d$ & Predicted embeddings from the GRU in the discriminator & 150 \\
        \bottomrule
    \end{tabular}
    }
    \vspace{-10pt}
\end{table}

\section{Dataset and Training}
We train our network on the TED Gesture Dataset~\cite{cospeech_gestures}, which consists of videos of English-language speakers at TED Talks. It provides 3D pose sequences of the upper-body gestures of the speakers, their speech audio, and the associated text transcripts. Each data sample has a sequence length of $T=34$ time steps at a rate of 15 fps. There are 200,038 training samples in total, constituting around 80\% of the dataset. The evaluation set consists of 26,903 samples or around 10\% of the dataset. The test set consists of 26,245 samples, making up the remaining 10\% of the dataset.

We use loss functions $\mathcal{L}_G$ and $\mathcal{L}_D$ identical to Yoon et al.~\cite{trimodal} to train our generator and discriminator respectively:
\begin{align}
    \mathcal{L}_G &= \lambda_{Hub}\mathcal{L}_{Hub} + \lambda_{gen}\mathcal{L}_{gen} + \lambda_{stl}\mathcal{L}_{stl} + \lambda_{KLD}\mathcal{L}_{KLD}, \\
    \mathcal{L}_D &= -\mathbb{E}\bracks{\log\parens{\textrm{Disc}\parens{U}}} - \mathbb{E}\bracks{\log\parens{1 - \textrm{Disc}\parens{\hat{U}}}},
\end{align}
where $\textrm{Disc}$ denotes the discriminator network (Sec.~\ref{subsec:discriminator}), $\lambda_*$ are the weights of the corresponding loss terms with the same values as in Yoon et al.~\cite{trimodal}, and the individual loss terms of the generator are:
\begin{itemize}
    \item \textbf{Huber loss}~\cite{huber_loss} between the ground truth and predicted edge vectors,
    \item \textbf{generative adversarial loss} on the output of the discriminator,
    \begin{equation}
        \mathcal{L}_{gen} = -\mathbb{E}\bracks{\log\parens{\textrm{Disc}\parens{\hat{U}}}},
    \end{equation}
    \item \textbf{diversity regularization} between the synthesized gestures and other gestures in the dataset to ensure that the styles of different speakers appear visually different,
    \item \textbf{Kullback-Leibler (KL) divergence} between the latent distribution space of the styles defined by $\mu_s$ and $\Sigma_s$, and the normal distribution $\mathcal{N}\parens{0, I}$.
\end{itemize}
Table~\ref{tab:hyperparams} lists the latent dimensions we use for training our network. We use the Adam optimizer~\cite{adam} with $\beta_1 = 0.5$, $\beta_2 = 0.999$, batch size of $512$, and learning rate of $5\textsc{e}^{-4}$ for the generator and $1\textsc{e}^{-4}$ for the discriminator with no warm-up epochs (\textit{i.e.}, $\lambda_{gen} > 0$ starting from the first epoch). We train our network for $300$ epochs, which took close to $45$ hours on an NVIDIA GeForce GTX 1080 Ti GPU.

\begin{table}[t]
    \centering
    \caption{Evaluation of our method with baselines and ablated versions of our method on two benchmark dataset, using the objective metrics of mean absolute joint error (MAJE), mean acceleration difference (MAD), and the Fr\'echet Gesture Distance (FGD). Bold indicates best. }
    \label{tab:objective_evaluation}
    \resizebox{\columnwidth}{!}{
    \begin{tabular}{clccc}
        \toprule
        Dataset & Method & MAJE (mm) & MAD (mm/s$^2$) & FGD \\
        \midrule
        \multirow{5}{2.4cm}{TED Gesture~\cite{cospeech_gestures}} & Seq2Seq~\cite{cospeech_gestures} & 45.62 & 6.33 & 6.62 \\
        & S2G-IS~\cite{individual_gesture_styles} & 45.11 & 7.22 & 6.73 \\
        & JEM~\cite{language2pose} & 48.56 & 4.31 & 5.88 \\
        & GTC~\cite{trimodal} & 27.30 & 3.20 & 4.49 \\
        \cmidrule{2-5}
        & Ours w/o MFCC Enc. & 27.84 & 3.02 & 4.21 \\
        & Ours w/o Aff. Enc. & 25.38 & 3.51 & 4.84 \\
        \cmidrule{2-5}
        & Ours & \textbf{24.49} & \textbf{2.93} & \textbf{3.54} \\
        \midrule
        \multirow{2}{2.4cm}{GENEA Challenge 2020~\cite{genea_2021}} & Gesticulator~\cite{gesticulator} & 82.41 & 3.62 & 31.04 \\
        \cmidrule{2-5}
        & Ours w/o MFCC Enc. & 105.71 & 1.57 & 23.03 \\
        & Ours w/o Aff. Enc. & 92.90 & 2.81 & 24.28 \\
        \cmidrule{2-5}
        & Ours & \textbf{54.93} & \textbf{1.49} & \textbf{20.36} \\
        \bottomrule
    \end{tabular}
    }
    \vspace{-10pt}
\end{table}

\begin{figure*}[t]
    \centering
    \includegraphics[width=\textwidth]{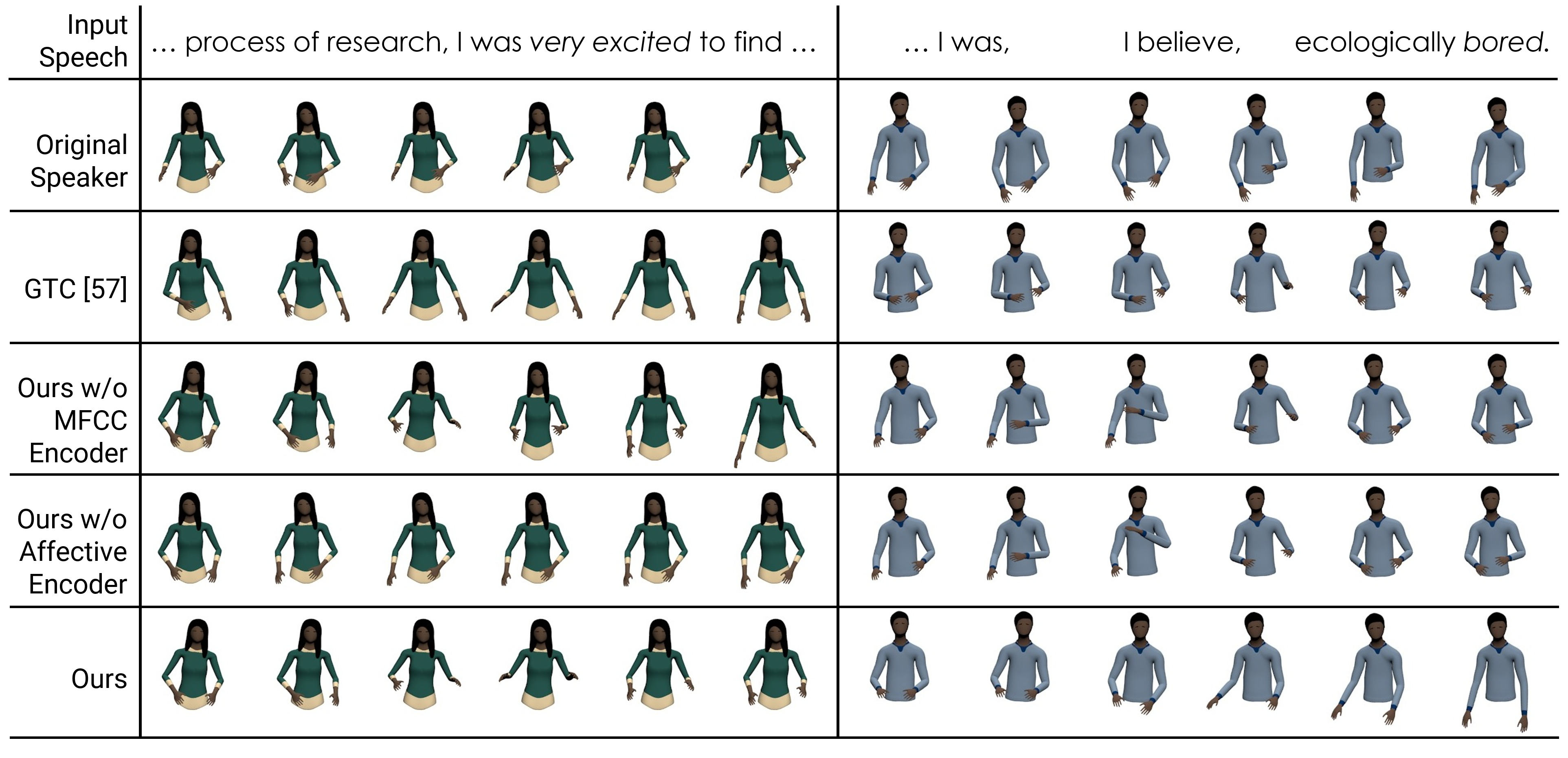}
    \caption{Qualitative results on the gestures synthesized by our method for two sample speech excerpts from the TED Gesture Dataset~\cite{cospeech_gestures}. The italicized words \textit{very excited} and \textit{bored} indicate the primary affect in the corresponding speeches. We compare with the corresponding gestures of the original speakers, the output of GTC~\cite{trimodal}, and that of the two ablated versions of our network (Sec.~\ref{subsec:ablation}). See Sec.~\ref{subsec:qualitative} for a detailed discussion of the results.}
    \label{fig:qualitative}
    \vspace{-10pt}
\end{figure*}

\section{Experiments}
We describe the objective evaluation of our method compared to current baseline methods. We highlight the benefit of our proposed components via ablation studies on the objective evaluation metrics. We also show the qualitative performance of our method on selected samples from the TED Gesture Dataset~\cite{cospeech_gestures} and the perceived quality of our synthesized gestures through a user study.
\subsection{Baseline Methods}
We compare our method with the baseline methods on two benchmark datasets, the TED Gesture Dataset~\cite{cospeech_gestures}, and the GENEA Challenge 2020 Dataset~\cite{genea_2020,genea_2021}.

On the TED Gesture Dataset, we compare with the methods of Seq2Seq~\cite{cospeech_gestures}, Speech to Gestures with Individual Styles (S2G-IS)~\cite{individual_gesture_styles}, Joint Embedding Model (JEM)~\cite{language2pose}, and Gestures from Trimodal Context (GTC)~\cite{trimodal}. Seq2Seq and JEM generate gestures based only on the text transcript of the speech, whereas S2G-IS uses only the speech to generate the gestures. GTC uses the speech, the corresponding text transcript, and the speaker styles to generate gestures. Seq2Seq follows an encoder-decoder architecture, where the authors transform the text to latent features and predict gestures based on both the latent features and a short gesture history. The authors of S2G-IS employ a generative adversarial network that generates gestures from a latent space obtained from the input log-Mel spectrograms. JEM maps both the text and the target gesture into a common latent embedding space and uses a decoder to reconstruct the gestures from the embedding space. The authors train the model to learn to align the text-based and the gesture-based embeddings for the same input and decode gestures from only the text-based embeddings. For Seq2Seq, S2G-IS, and JEM, we follow the training routine and the hyperparameters used by Yoon et al.~\cite{trimodal}. For GTC, we directly use the pre-trained model provided by Yoon et al.~\cite{trimodal}.

The GENEA Challenge 2020 Dataset is the publicly available version of the Trinity Gesture Dataset~\cite{genea_2020,genea_2021}. It consists of the speech and the full-body motion capture of a male actor talking unrestrained about various topics over multiple recording sessions. The full dataset is about 242 minutes long, of which 221 minutes are used as training data, and the remaining 21 minutes are kept for testing. We do not fine-tune our network on this dataset and evaluate our network on the test partition. Since we consider only upper-body gestures, we consider the ten relevant upper-body joints at the root, the spine, the head, and the two arms for evaluating our performance. On this dataset, we compare with the method of Gesticulator~\cite{gesticulator}, which leverages the acoustics and the semantics of the speech to generate semantically consistent beat, deictic, metaphoric, and iconic gestures. For a fair comparison, we use the pre-trained model provided by the authors and compare the performance on the same ten joints that we use for our method.

\subsection{Objective Evaluation}
While evaluation metrics for gesture synthesis are not standardized, we evaluate on the commonly used metrics of mean absolute joint error (MAJE), mean acceleration difference (MAD), and the Fr\'echet Gesture Distance (FGD) proposed by Yoon et al.~\cite{trimodal}. MAJE measures the mean of the absolute differences between the ground truth and the predicted joint positions over all the time steps, joints, and samples. MAD measures the mean $\ell_2$-norm error between the ground truth and predicted joint accelerations over all the time steps, joints, and samples. FGD measures the difference between the distributions of the latent features of the ground truth and the predicted gestures. The latent features are computed from an autoencoder network trained on the well-known Human 3.6M dataset~\cite{human36m} of human motions using the ten joints in the TED Gesture Dataset~\cite{cospeech_gestures}. MAJE indicates how closely the predicted joint positions follow the ground truth joint positions. MAD indicates how closely the ground truth and predicted joint movements match. Since affective expressions are based on joint movements, a lower MAD is especially desirable for our stated aim of generating gestures with appropriate affective expressions. FGD is shown to align well with the perceived plausibility of the synthesized gestures to human users~\cite{trimodal}; therefore, a lower FGD is equally desirable to gauge the quality of our synthesized gestures.

Table~\ref{tab:objective_evaluation} summarizes the performance of all the methods on all these evaluation metrics. Our method consistently has the lowest MAJE, MAD, and FGD on both the benchmark datasets. On the TED Gesture Dataset, we observe improvements of 10.29\%, 8.44\%, and 21.16\% on MAJE, MAD, and FGD, respectively, over the best current baseline of GTC. On the GENEA Challenge 2020 Dataset, we observe improvements of 33.34\%, 58.84\%, and 34.41\% on MAJE, MAD, and FGD, respectively, over the baseline of Gesticulator. We note that the absolute FGD values are significantly higher on the GENEA Challenge 2020 Dataset than on the TED Gesture Dataset. We hypothesize that this is because the gestures in the GENEA Challenge 2020 Dataset are more abstract and unscripted compared to the well-defined actions in the Human 3.6M Dataset or polished speeches in the TED Gesture Dataset. As a result, the pre-trained latent embeddings used for FGD are not as good at reconstructing the joints movements in the GENEA Challenge 2020 Dataset.

\subsection{Ablation Studies}\label{subsec:ablation}
We perform ablation studies on our two proposed components: the MFCC encoder (Sec.~\ref{subsubsec:mfcc_encoder}) and the affective encoder (Sec.~\ref{subsubsec:aff_encoder}). In one study, we replace only our MFCC encoder with an encoder for the raw audio waveform, identical to that of GTC~\cite{trimodal}, and train the resultant network. In the other study, we remove the affective encoder from our network. Our generator takes in the raw seed poses instead of the latent affective features. Our discriminator uses a convolution filter identical to GTC~\cite{trimodal} to transform the input gestures to latent features for the bidirectional GRU.

Without the MFCC encoder, our generator cannot take the speech-based affective cues into account. It results in a degradation of the synthesis, leading to higher MAJE, MAD, and FGD. However, without the affective encoder, our network is severely limited in understanding both the affective expressions in the seed poses and the affective expressions of the synthesized poses. It results in more severe performance degradation on all the evaluation metrics, and the synthesized gestures appear less diverse and plausible.

\begin{figure}[t]
    \centering
    \begin{subfigure}[b]{\columnwidth}
        \centering
        \includegraphics[width=\textwidth]{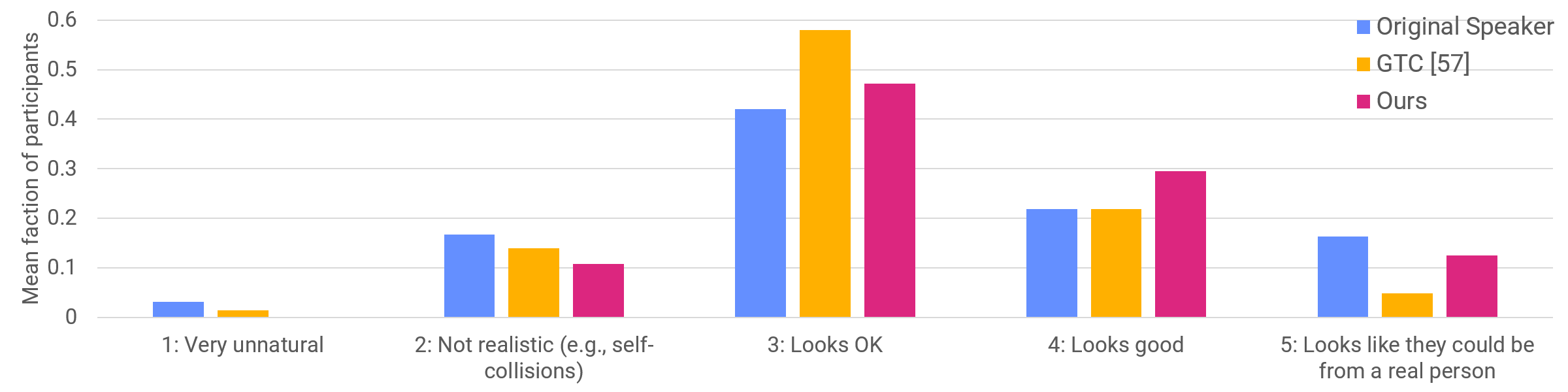}
        \caption{Plausibility of the different types of gestures.}
        \label{fig:plausibility_likert_scale}
    \end{subfigure}
    \begin{subfigure}[b]{\columnwidth}
        \centering
        \includegraphics[width=\textwidth]{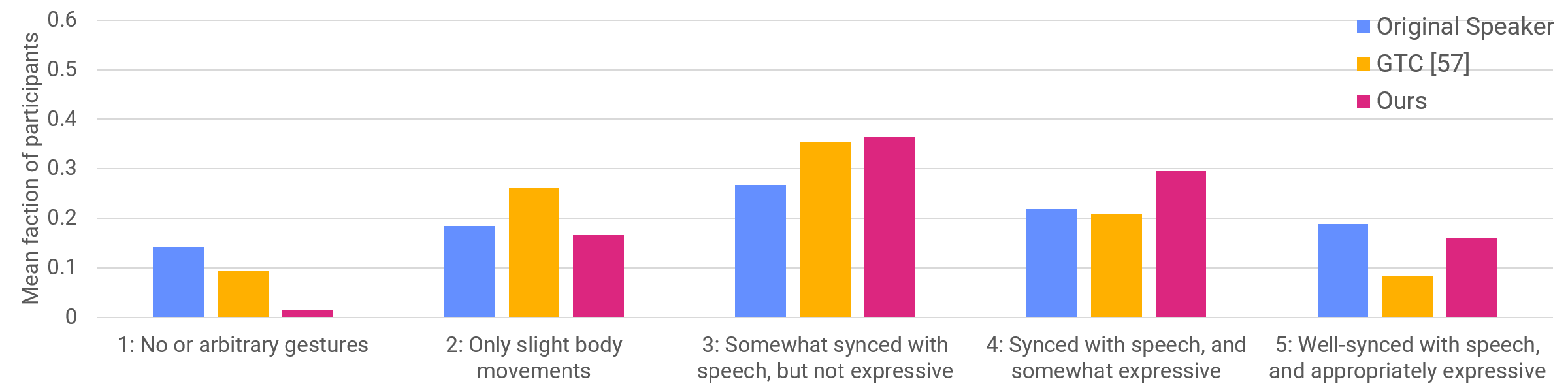}
        \caption{Synchronization of the movements and the affective expressions of the different types of gestures with the speech.}
        \label{fig:sync_to_speech_likert_scale}
        \vspace{-7pt}
    \end{subfigure}
    \caption{Mean fraction of participant responses on each point of the Likert scales across the 12 speech excerpts from the TED Gesture Dataset~\cite{cospeech_gestures} and the corresponding gestures in our user study. See Sec.~\ref{subsec:user_study} for details.}
    \label{fig:user_study}
    \vspace{-10pt}
\end{figure}

\subsection{Qualitative Results}\label{subsec:qualitative}
We show qualitative results on two sample speech excerpts from the TED Gesture Dataset~\cite{cospeech_gestures} in Fig.~\ref{fig:qualitative}. It has five rows of gestures: those of the original speakers', those synthesized by GTC~\cite{trimodal} ( the current \sota), those by the two ablated versions of our network: one without our MFCC encoder (Sec.~\ref{subsubsec:mfcc_encoder}) and the other without our affective encoder (Sec.~\ref{subsubsec:aff_encoder}), and those by our proposed network with all the encoders. We observe a diversity of speaker styles in the synthesized gestures compared to the original speaker, which results from using a variational embedding of speaker styles using the speaker encoder (Sec.~\ref{subsubsec:speaker_encoder}). GTC, however, cannot generate affective expressions except for a few words with strong intonations in the speech, such as ``excited'' (second row, left column). Without our MFCC encoder, our network can still match the speech content but cannot align the gestures with the affective cues from the speech. For example, it can match the words ``I was, I believe'' with a deictic gesture pointing to the speaker himself (third row, right column) but cannot generate any expressions for ``bored''. Without our affective encoder, we observe only slight body movements but no appreciable affective expressions in the synthesized gestures. With all our encoders in place, we observe appropriate affective expressions that align well with the speech. For example, we observe rapid arm movements when saying ``excited'' (fifth row, left column) and dropping of the arms and shoulders when saying ``bored'' (fifth row, right column).

\subsection{User Study}\label{subsec:user_study}
We conducted a user study to evaluate the perceptual quality of our synthesized gestures in terms of how plausible they appear and how well-aligned are their affective expressions with the corresponding speeches. 24 participants took part in our study, of which 20 were male, and 4 were female. 10 participants were between 18 and 24 years of age, 13 were between 25 and 34, and one was above 35. Each participant observed gestures corresponding to the same 12 speech excerpts, each taken from a different TED Talk in the TED Gesture Dataset~\cite{cospeech_gestures}. For each speech excerpt, the participants observed three different types of gestures: that of the original speaker as a 3D pose sequence (provided in the dataset), those synthesized by GTC~\cite{trimodal}, the current \sota, and those synthesized by our network. The order of the gestures was unknown to and randomized for each participant. We then asked the participants to answer two questions. The first question was how plausible the gestures appeared on a five-point Likert scale ranging from ``very unnatural'' (1) to ``look like they could be from a real person'' (5). The second question was how well the gestures synchronized with the corresponding speeches on a five-point Likert scale, ranging from ``no or arbitrary gestures'' (1) to ``well-synchronized with the speech, and are appropriately emotionally expressive'' (5). Intuitively, our Likert-scale points for both questions reflect the participants' individual assessments of quality, with 1 being the worst, 3 being average, and 5 being the best. The entire study took around 20 minutes on average for each participant.

We summarize the participants' responses in Fig.~\ref{fig:user_study}. When adjudging the plausibility of the gestures (Fig.~\ref{fig:plausibility_likert_scale}), we observe that 15.28\% more participants marked our synthesized gestures either 4 or 5 compared to the gestures synthesized by GTC~\cite{trimodal}. Further, 3.82\% more participants marked our synthesized gestures 4 or 5 than the original speakers' gestures, indicating that the participants found our synthesized gestures to have visual quality comparable to that of the original data.
When adjudging the synchronization of the movements and the affective expressions of different types of gestures with speech (Fig.~\ref{fig:sync_to_speech_likert_scale}), we observed that 16.32\% more participants marked our synchronization quality either 4 or 5 compared to that of GTC~\cite{trimodal}. Also, 4.86\% more participants marked out synchronization quality 4 or 5 than that of the original speakers, indicating that the participants perceived our synthesized gestures to be as well-synchronized and expressive as the original data.

\section{Conclusion, Limitations and Future Work}
We have presented an end-to-end learning approach to generate 3D pose sequences of co-speech gestures with appropriate affective expressions. Our contributions include an MFCC encoder to guide the gesture synthesis based on the speech-based affective cues such as prosody and intonation, and an affective encoder to learn joint-based affective features from the gesture data. Using these encoders in a generative adversarial learning framework, we have synthesized affective gestures that advance the state-of-the-art on co-speech gesture synthesis on multiple evaluation metrics. Our synthesized gestures also appeared more plausible and well-synced with the corresponding speeches to participants in a user study.

Our work has some limitations. First, we do not build a mechanism to control the affective expressions; we compute them automatically from the input modalities. We plan to investigate the interface between affective expressions from the speech and the gestures, especially when expressing contradictory cues such as sarcasm and irony. Second, we plan to use a finer representation of poses in the future since affective expressions in the gestures are often associated with subtle movements not captured by our current representation. Lastly, our network uses only upper-body gestures with a fixed root. We plan to expand to affect-aware gestures of the whole body, incorporating locomotion in the global coordinate space. We also plan to align the body gestures with the corresponding facial expressions, leading to fully emotive characters.


\section*{Acknowledgment}
This work has been supported in part by ARO Grants W911NF1910069 and W911NF1910315, and Intel.



\bibliographystyle{IEEEtran}
\bibliography{arxiv}
%



\end{document}